# The game of metaphysics: towards a fictionalist (meta)metaphysics of science

Raoni Arroyo

Centre for Logic, Epistemology and the History of Science

University of Campinas

Matteo Morganti

Department of Philosophy, Communication, and Performing Arts

Roma Tre University

***RÉSUMÉ:*** *La métaphysique est traditionnellement conçue comme visant la vérité — en réalité les vérités les plus fondamentales sur les caractéristiques les plus générales de la réalité. Les partisans du naturalisme philosophique, qui insistent pour que les revendications philosophiques soient fondées sur la science, ont souvent adopté une attitude éliminativiste à l'égard de la métaphysique, n'accordant par conséquent que peu d'attention à cette définition. Dans la littérature plus récente, toutefois, le naturalisme a plutôt été interprété comme signifiant que la conception traditionnelle de la métaphysique ne peut être acceptée que si l'on est réaliste scientifique (et que l'on met les bonnes contraintes sur les revendications métaphysiques acceptables). Nous voulons suggérer ici que les naturalistes peuvent, et peut-être devraient, choisir une troisième option, fondée sur une révision significative mais acceptable de la compréhension établie de la métaphysique. Plus précisément, nous affirmerons qu'une approche fictionnaliste de la métaphysique est compatible à la fois avec l'idée que la discipline s'intéresse aux caractéristiques fondamentales de la réalité et avec la méthodologie naturaliste; elle s'accorde bien à la fois avec le réalisme scientifique et avec l'instrumentalisme.*
***Mots clés:*** *métaphysique; naturalisme; fictionnalisme; (anti-)réalisme; underdétermination.*

***ABSTRACT:*** *Metaphysics is traditionally conceived as aiming at the truth — indeed, the most fundamental truths about the most general features of reality. Philosophical naturalists, urging that philosophical claims be grounded on science, have often assumed an eliminativist attitude towards metaphysics, consequently paying little attention to such a definition. In the more recent literature, however, naturalism has instead been taken to entail that the traditional conception of metaphysics can be accepted if and only if one is a scientific realist (and puts the right constraints on acceptable metaphysical claims). Here, we want to suggest that naturalists can, and perhaps should, pick a third option, based on a significant yet acceptable revision of the established understanding of metaphysics. More particularly, we will claim that a fictionalist approach to metaphysics is compatible with both the idea that the discipline inquires into the fundamental features of reality and naturalistic methodology; at the same time, it meshes well with both scientific realism and instrumentalism.*

## Metaphysics, science and truth

How do contemporary metaphysicians perceive the overarching purpose of metaphysics? A glance at McKenzie's[1] recent survey of the literature provides a good answer:

---

[1] K. McKenzie, *Fundamentality and Grounding*, Cambridge, Cambridge University Press, 2022.

"Metaphysics has traditionally been thought of as the systematic study of the most fundamental structure of reality — and indeed, this is the view of it that I would like to support.";[2] "[…] [Metaphysics] is about the structure of the world.";[3] "The heart of metaphysics is the question: what is the world ultimately, or fundamentally, like?";[4] "Metaphysics concerns the search for general and fundamental truths about the world. […] [T]he metaphysician should be concerned to prescriptively develop and understand the prior, deep, and general truths about the fundamental natures of the world.";[5] "In a nutshell, metaphysics is the study of the fundamental structure of reality".[6]

McKenzie brings up these quotations as she takes them to be revealing of the discipline's goal, *i.e.*, answering fundamentality questions. This won't be questioned here, as our focus will be on another aspect of these passages which we take to be equally revealing: all of the above authors represent metaphysics as aiming for a true description of the world, *i.e.*, as seeking (fundamental) truths about reality. Here, we will refer to this as 'realism about metaphysics'.

A well-documented criticism of this goal is epistemic, *viz.*, how exactly can one know whether metaphysics has delivered the goods of truly describing (fundamental) reality? Some suggest that metaphysics has privileged access to fundamental truths thanks to its *a priori* methodology and, possibly, the fact that it employs sui generis, peculiar faculties — the label 'rational intuition' is often used.[7] We will not discuss this perspective in this paper, if only because it is not very popular nowadays, and will instead tackle the issue from a different angle. Plausible objections to this approach include the implausibility of the view that rational intuition constitutes a sufficient a source of justified belief on its own; the idea that logic and conceptual analysis alone hardly suffices to single out truths about reality; and, more importantly for present purposes, that it seems odd to regard metaphysics as provided with a potential to provide us with knowledge that even science — no doubt the paradigm of a successful epistemic endeavour — lacks.

The problem has to do with an alleged lack of evidential support. Since it is detached from the empirical domain, an often-heard argument goes, metaphysics doesn't get the systematic support from empirical evidence that, most notably, science has. Indeed, metaphysics is often seen as completely detached from science and, because of this, it is criticised for being in a weak spot, epistemologically speaking. Whereas science succeeds in its explanatory endeavours thanks to its distinctive methodology, and in particular its peculiar anchoring in empirical data and experimentation, it is not so in the case of metaphysics. The ensuing difficulty of showing exactly where the relevance of metaphysics lies, and what would ground the preference for a metaphysical hypothesis over an alternative, has often led to scepticism towards the enterprise as a whole, and to requests for its complete dissolution.[8]

---

[2] J. Lowe, *The possibility of Metaphysics: Substance, Identity, and Time*, Oxford, Clarendon Press, 1998.
[3] J. Schaffer, "On What Grounds What", *in* D. Chalmers, D. Manley, R. Wasserman (eds.), *Metametaphysics: New Essays on the Foundations of Ontology*, Oxford, Oxford University Press, 2009, pp. 347–83.
[4] T. Sider, *Writing the Book of the World*, Oxford, Oxford University Press, 2011.
[5] L. A. Paul, "Metaphysics as Modeling: The Handmaiden's Tale", *Philosophical Studies*, 160(1), 2012, pp. 4–6.
[6] T. Bigaj, C. Wüthrich, "Introduction", *in* T. Bigaj & C. Wüthrich (eds.), *Metaphysics in Contemporary Physics*, Rodolpi, Brill, 2016, p. 8.
[7] See, for example, G. Bealer, "Intuition and the Autonomy of Philosophy", *in* M. DePaul & W. Ramsey (eds.), *Rethinking Intuition: The Psychology of Intuition and Its Role in Philosophical Inquiry*, Rowman and Littlefield, 1998, pp. 201–40.

A growingly popular remedy to this supposed deficiency of metaphysics consists in establishing a tight connection of metaphysics with science. So-called 'naturalists', in particular, recommend that metaphysics be made as continuous as possible with science at the level of ontological commitment and/or methodology. This may be achieved, among other things, by only including in one's ontology the sort of entities and processes that are postulated by our best scientific theories; or by employing in metaphysics the same tools that seem to be used in science, starting from empirical observation and building explanatory models on the basis of something like inference to the best explanation. Obviously enough, this kind of naturalism is quite different from the 'eliminative' naturalism that is sometimes put forward on the basis of philosophical assumptions that ultimately lead back to neopositivism — according to which the success of science proves metaphysical claims and hypotheses to be meaningless or at any rate dispensable. That metaphysics should be continuous with science has lately become a common belief, as witnessed by numerous contributions and debates in metametaphysics.[9] Naturalised, non-eliminative metaphysics thus allegedly fixes that problem with the epistemic credibility of metaphysics.

What remains under dispute is the amount of traditional metaphysics that can be salvaged with such a manoeuvre. As Chakravartty put it, the goals of naturalisation can be cashed out in different terms:

> It is not uncommon to hear that continuity in this context is evidenced by the fact that naturalized metaphysics is 'derived from', 'based on', or otherwise 'inspired' or 'motivated' or 'constrained by' our best science, which thereby serves as the proper 'ground' for metaphysical theorizing.[10]

These terms suggest slightly different views of naturalised metaphysics. One useful way of understanding this is in terms of epistemic warrants. While logical empiricists regarded metaphysical claims as literally meaningless due to the role they attributed to a criterion of verifiability, contemporary philosophers in the empiricist tradition do not dispute the meaningfulness of metaphysical debates. Rather, they urge the naturalisation of metaphysics in the sense that only hypotheses and claims with a sufficient degree of empirical support should be taken seriously. What exactly should be deemed 'sufficient' is, then, open to debate. Chakravartty[11] convincingly argues that the choice appears to be ultimately a matter of personal preference, as it has to do with what balance one finds most congenial between a) the 'epistemic safety' provided by empirical support, and b) the potential gains in terms of knowledge, explanation and/or understanding provided by 'more daring' hypotheses.

---

[8] B. van Fraassen, *Quantum Mechanics: An Empiricist View*, Oxford, Oxford University Press, 1991; H. Putnam, *Ethics without Ontology*, Cambridge, MA, Harvard University Press, 2004; J. Ritchie, *Understanding Naturalism*, Durham, Acumen, 2008; D. Ross, "Vikings or Normans? The Radicalism of Naturalized Metaphysics", *Metaphysica*, 17(2), 2016, pp. 213–27; R. Arroyo, J. R. B. Arenhart, D. Krause, "The Elimination of Metaphysics through the Epistemological Analysis: Lessons (un)Learned from Metaphysical Underdetermination", in D. Aerts et al. (eds.), *Probing the Meaning of Quantum Mechanics: Probability, Metaphysics, Explanation and Measurement*, Singapore, World Scientific, 2023, pp. 278–324.

[9] For panoramic surveys, see J. R. B. Arenhart, R. Arroyo, "The Spectrum of Metametaphysics: Mapping the State of Art in Scientific Metaphysics", *Veritas*, 66(1), p. e41217, 2021, and M. Morganti, *Metaphysics and the Sciences*, Cambridge, Cambridge University Press, 2024; for a recent original proposal, see N. Emery, *Naturalism Beyond the Limits of Science: How Scientific Methodology Can and Should Shape Philosophical Theorizing*, New York, Oxford University Press, 2023.

[10] A. Chakravartty, "On the Prospects of Naturalized Metaphysics", in D. Ross, J. Ladyman, & H. Kincaid (eds.), *Scientific Metaphysics*, Oxford, Oxford University Press, 2013, pp. 40–1.

These questions, not surprisingly, also concern one's views on the epistemic import of science, *i.e.*, one's position with respect to the issue of scientific realism. There seem to be two important conditional assumptions at play in most, if not all, recent discussions of naturalism and realism. The first is the following:

(1) Naturalism → (Scientific antirealism → antirealism in metaphysics)

Scientific antirealists deny that *e.g.*, quarks truly inhabit the world in which we live: at best, they inhabit *our (best) scientific theories*. When science talks about such things, it does so in a merely pragmatic way, as a placeholder for "*whatever a quark does*". Indeed, says the antirealists, particle physics works properly independently of whether one believes that quarks 'really' exist. As the very subject matter of metaphysics is a domain of unobservable entities and processes, it consequently seems natural for scientific antirealists to extend the same attitude to it as well. As Hawley put it,

> […] it should come as no surprise that anyone who is sceptical about the ability of science to give us knowledge of quarks and quasars will be sceptical about whether science can give us knowledge of universals and possible worlds.[12]

Adding to this the thought that 'metaphysical unobservables' are hardly of any use when it comes to saving the phenomena — a task for which the entities posited by scientists are amply sufficient — it can be easily seen why scientific antirealism is normally accompanied by metaphysical eliminativism.[13] The second common assumption is the following bi-conditional:

(2) Naturalism → (Scientific realism ↔ realism in metaphysics)

In Ladyman's words, "[a]nalytic metaphysics and science both seek general truths about reality."[14] Obviously enough, not every piece of metaphysics will do: let us recall

---

[11] A. CHAKRAVARTTY, *Scientific Ontology: Integrating Naturalized Metaphysics and Voluntarist Epistemology*, Oxford, Oxford University Press, 2017.
[12] K. HAWLEY, "Science as a Guide to Metaphysics?", *Synthese*, 149(3), 2006, pp. 451–70.
[13] Here we are of course ignoring the possibility of being an antirealist in science and a realist in metaphysics, as that would not qualify as a naturalistic attitude towards metaphysics.
[14] J. LADYMAN, "An Apology for Naturalized Metaphysics", *in* M. SLATER, Z. YUDELL (eds.), *Metaphysics and the Philosophy of Science: New Essays*, Oxford, Oxford University Press, 2017, pp. 141–162.

Ladyman and Ross's opening of their famous manifesto, where they called for the reform of metaphysics: non-naturalistic metaphysics, in their analysis, "fails to qualify as part of the enlightened pursuit of objective *truth*".[15] Rather, metaphysical claims are warranted only to the extent that they respect certain constraints related to scientific theorising. In Ladyman and Ross's case, metaphysical hypotheses are to be accepted only if they serve to unify distinct parts of science. Other authors are more liberal and just require that metaphysical hypotheses be relevant for the interpretation of scientific theories. Against this background, assuming a realist picture of science the naturalist is led to believe that (aptly constrained) metaphysics inherits the same epistemic status. On the other hand, as we have just indicated, in the naturalistic context it is commonly believed that the only way to be a non-eliminativist about metaphysics is to be a realist about it. And, since realism about metaphysics without scientific realism is not an option for the naturalist, (2) follows.

A sort of 'epistemic parity' assumption seems to emerge: either antirealism towards both science and metaphysics; or realism towards both science and in metaphysics. Such a thesis, even though not under this name and with some qualifications, is explicitly put forward by Emery, who states:

> If there is a mismatch between one's view about science and one's view about metaphysics — if, for instance, one is a pragmatist about metaphysics but a realist about science, or if one is a realist about metaphysics but a pragmatist about science — then it would be odd to be a […] naturalist. After all, someone who endorses this kind of mismatch thinks that science and metaphysics have significantly different goals — so why should they care if their scientific claims and their metaphysical claims conflict.[16]

We will get back to this later. In what follows, we will question this *'parity thesis'*, and argue that naturalists should just, more weakly, require our epistemic attitude towards metaphysics *not to be stronger* than the attitude we have towards science. In particular, if one were to show that some form of non-realist metaphysics can be consistently endorsed along with scientific realism in a naturalistic context, the implication in (2) would fail (and, obviously enough, the reverse of (1) would not hold either). In the final part of the paper, we will defend exactly this view, suggesting that it represents the best way of putting non-eliminative naturalism about metaphysics into practice.

# Fictionalist metametaphysics

Our starting point is the common thought that, for a scientific antirealist, naturalism about metaphysics means elimination (inference (1) above). A remarkable recent exception to this widespread belief is represented by Emery.[17] She argues that, as far as the naturalist enterprise is concerned, the antirealist stance in science is perfectly compatible with a non-eliminativist antirealist stance in metaphysics. The metaphysical naturalist position, recall, is broadly understood as the methodological continuity between science and metaphysics. Since the naturalist establishes such a complete methodological continuity, Emery argues, she can be a 'pragmatist' (Emery's terminology) about scientific unobservables as much as about metaphysical unobservables, attributing the same non-truth-related epistemic role (*e.g.*, providing satisfactory explanations, increasing understanding *etc*.) to both. We completely agree with Emery's claim that an instrumentalist/pragmatist attitude towards metaphysics is viable — we will get back to this later. However, for reasons that we will present in a moment, we disagree with her views on the actual amount of continuity that can be found between science and metaphysics. As we will see, Emery's assumption led her to unwarrantedly endorse the parity thesis.

---

[15] J. LADYMAN, and D. ROSS, *Every Thing Must Go: Metaphysics Naturalized*, Oxford, Oxford University Press, 2007, p. vii, emphasis added.
[16] N. EMERY, *Naturalism Beyond the Limits of Science: How Scientific Methodology Can and Should Shape Philosophical Theorizing*, New York, Oxford University Press, 2023, p. 30.
[17] *Ibid.*, pp. 54–5.

To begin our discussion, it is essential to spell out pragmatism/antirealism about metaphysics in some detail. Obviously enough, going along this route requires rejecting the idea that the respectability of the metaphysical enterprise rests on the truth of its claims, *viz.*, on the possibility of interpreting them literally, in terms of correspondence with the world. But how can this possibly be done given that, as we have seen, metaphysics is normally *defined* as the search for fundamental truths about reality?

We believe that it is essential to distinguish the role that truth plays when *doing* metaphysics and when *talking about* it, in particular when assessing its epistemic import, respectively. As we see it, it is perfectly legitimate to explore different metaphysical hypotheses and theories and take each one of them seriously (even so much as to 'play the truth game') only as long as one works with it. Whether a particular metaphysical hypothesis is objectively true, however, can be regarded as something like a Carnapian external question that can — and perhaps must — be left unanswered. We are using Carnap's distinction between external and internal questions for illustration purposes only. The view of metaphysics that we will put forward does not require it, as it is essentially a view about epistemic attitudes, not about the context-dependence of meaningfulness and/or the impossibility to get outside of a particular framework. Even less do we need to endorse the sort of syntactic view of scientific theories assumed by Carnap's notion of a linguistic framework. After all, something similar happens, one may plausibly conjecture, in the scientific context: surely, a physicist working with a particle accelerator believes that there are particles inside the machine while an experiment is being performed. And they can also believe that, more generally, physics is correctly described as the search for knowledge of fundamental truths about reality. At the same time, though, they may well abstain from answering 'yes' when asked 'Do those particles *really* exist (as described by the theory that you use to design and conduct your experiments)?' when they come out of the lab. Such a switch from belief to acceptance, we think, is by no means inconsistent or delusional. To the contrary, it could be argued that this sort of 'ambiguity' underlies a lot of what goes on in current analytic metaphysics and metaphysics of science, and perhaps even in our day-to-day experience. Are we assuming a realist commitment, for instance, when we say 'I like the number 7 more than I like the number 4', or 'Inflation has caused many to lose their jobs'?

This view can be regarded, with Emery, as a form of pragmatism or, alternatively, as the sort of empiricist instrumentalism according to which the function of our hypotheses about the unobservable is to 'save the phenomena'. In the case of metaphysics, one would account for the phenomena by providing useful explanations of them based on a peculiar vocabulary and typical conceptual tools and categories, different from those of science. Another option would be to endorse Bueno's 'neo-Phyrronism'. According to the neo-Pyrrhonist, even if truth is not the aim of inquiry, it still makes sense to construct hypotheses based on unobservable entities, mechanisms and processes of the metaphysical kind. For, doing so one might "[…] obtain an understanding of the various possibilities that are available to make sense of the issues under consideration and the insights such possibilities offer […]".[18] Similarly, one could subscribe to the view of metaphysics as model-building put forward by *e.g.*, Godfrey-Smith,[19] Guay and Pradeu,[20] and Paul,[21] according to which, once again, the primary function of models is not to provide true descriptions of reality but rather to increase our understanding.

However, besides more specific worries that we don't need to, and cannot, get into here, we believe that all these options are ultimately not entirely satisfactory, and for the same reason. Namely, because they explicitly do away with the notion of truth altogether, at least as long as the unobservable is concerned, even at the level of the actual practice of the inquirer. We believe that this is a mistake: as we already explained, truth is an integral part of the game one plays when one inquires into the nature of things (be it through science or metaphysics). In this sense the traditional conception of metaphysics can and should be preserved — if only because we normally believe and act as if we are dealing with aspects of reality when we seek certain explanations. The key point is that a differentiation can be consistently made, we think, between the role that truth plays when one is in the process of performing the inquiry, and when one looks at the inquiry from the outside, as it were.

For this reason, we find it most illuminating to regard our suggested approach to metaphysics as a form of *fictionalism*, along lines first suggested by Rosen. As he puts it:

> […] just as it is reasonable for the fictionalists […] to go in for quantum chemistry even though they view the enterprises not as a search for truth but as an exercise in model-building, so it is reasonable for those of us with a taste for metaphysics to pursue metaphysics in a similar spirit.[22]

This to say that creating models in order to achieve something different from objective

---

truth is perfectly acceptable both in science and in metaphysics. In both cases we can use theories and evaluate them 'as if' the world were as they claim it to be. However, and this is the key point, the metaphysical fictionalist does not just accept a theory in the empiricist sense of believing it to be true as long as observable phenomena are concerned, and merely useful otherwise. In the (fictional) fictionalist's own explanation of their position:

> When we think we've found a theory that will survive as part of this ideal model [a general worldview that satisfies certain cognitive requirements], we accept it. We come to believe that it's acceptable — that's the cognitive element in acceptance. But we also make it ours, immersing ourselves in its world-picture, resolving to speak as if we believed it (Monday through Friday) [that is, when actually using the model], to rely on it for practical purposes even when the stakes are high, and to treat it as the basis for subsequent inquiry.[23]

We think that Rosen's depiction of fictionalism makes perfect sense, and in fact constitutes a compelling reconstruction of metaphysical practice. In particular, practising (scientists and) metaphysicians do not merely 'speak as if they believed' in the sense of consciously pretending: rather, they switch to an altogether different epistemic attitude when they 'immerse themselves in a particular world-picture'. While we cannot provide a fully specific version of fictionalism here, a few considerations can be usefully added to the broad characterisation of the view just provided.[24]

First of all, we believe that what is at stake in the present context is a form of fictionalism involving the *use* of certain expressions and the ideal *attitude* of the speakers using them, not the *meaning* of those expressions. For example, fictionalism about electrons or tropes is *not* intended here as the claim that electrons or tropes are fictional objects; rather, it means that someone who utters or hears 'Properties are tropes' or 'This is an electron' should a) take these sentences literally, b) believe that these sentences are true and not simply useful, empirically adequate *etc.*, c) be ready to give up such belief, or at least suspend judgement on it, in certain contexts — in particular, when moving to the meta-level, at which questions are asked about one's theories.

Another relevant point is that we do not subscribe to Rosen's differentiation between *scientific* and *speculative* metaphysics.[25] According to Rosen, the former is firmly grounded in science and tackles questions that scientists themselves 'may leave dangling'. The latter is instead relatively autonomous and answers sui generis non-scientific questions. Because of this, Rosen believes that only the latter requires the more cautious epistemic attitude represented by fictionalism. We don't think that this is the case. The reason is that, as soon as one asks a metaphysical question — albeit about, or inspired directly by, a scientific theory — one has to deal with an amount of underdetermination that makes a 'suspension of judgement' with respect to the issue of Truth advisable. We will lend support to this claim in the final section, where we consider a case study concerned with something that Rosen would regard as scientific metaphysics: i.e., the measurement problem in quantum mechanics and the related issue of interpretation.

---

[23] *Ibid.*, p. 30.
[24] 'Fictionalism' is definitely a loaded term that can be intended in multiple ways and has uses in several areas. For instance, one may distinguish between scientific, metaphysical, mathematical, moral, modal fictionalism, and so on. For general surveys, see M. Eklund, "Fictionalism", in E. Zalta, U. Nodelman (eds.), *The Stanford Encyclopedia of Philosophy*, Spring 2024 Edition, Metaphysics Research Lab, Stanford University, 2024, url: https://plato.stanford.edu/archives/spr2024/entries/fictionalism/, and N. Gentile, S. Lucero, "On Compatibility between Realism and Fictionalism: A Response to Suárez' Proposal", *Studies in History and Philosophy of Science*, 103, 2024, pp. 169 *ff.*
[25] G. Rosen, "Metaphysics as a Fiction", in B. Armour-Garb, F. Kroon (eds.), *Fictionalism in Philosophy*, New York, Oxford University Press, 2020, p. 28–47, p. 35–6.

Let us now move to the other main claim that we want to make in the present paper: that, contrary (2) above, the naturalist who is a scientific realist need not be a realist about metaphysics as well. As we have seen, scientific realists working in a naturalist setting have questioned the scope and autonomy of traditional metaphysics, but not its customary conception as the search for fundamental truths. Working under the (at least implicit) assumption that we called 'parity thesis', that is, naturalistically-inclined scientific realists have mostly taken for granted that a properly constrained metaphysics inherits the epistemic credibility of the science that underlies it.

When one looks for an explicit argument, however, things turn out to be far from straightforward.[26] To illustrate, we might consider the following two points of tension. First, science and metaphysics don't need to share the same epistemic goals for naturalism to hold — this isn't a necessary condition for naturalism. The naturalist demand for continuity between the two disciplines is entirely compatible with the view that science retains a privileged epistemic position in comparison to metaphysics. After all, virtually no naturalist denies that there is a difference between the two disciplines, as metaphysics remains further remote from the empirical data, as it were. Secondly, and more importantly for the present discussion, that the methodological continuity between science and metaphysics warrants the extension to the latter of one's realist attitude (in case one has it) towards the former is far from obvious. To the contrary, it can only appear compelling if one believes that one's epistemic attitude towards science depends on the methodology it employs. Indeed, this seems to be the basic idea behind the parity thesis: if I am an (anti)realist about science — the typical piece of reasoning seems to be — this is based on the features of scientific methodology; since naturalism amounts to establishing methodological continuity between science and metaphysics, then I should also be an (anti)realist about metaphysics. This line of thought, however, is incorrect. For, the best actual reasons for adopting realism towards science just do not have to do with the methodology that led scientists to formulate certain theories!

In particular, empirical data are clearly insufficient for picking out one particular scientific theory out of the many possible empirically adequate ones. Hence, theoretical virtues inevitably play a role in scientific theory construction and theory choice. However, even granting that theoretical virtues constitute an uncontroversial guide towards the 'best' theory in some sense of the term — which is quite controversial! —, this is still insufficient for scientific realism, for the simple reason that such virtues are commonly regarded as not being truth-conducive. Indeed, many false scientific theories of the past are likely to have been formed based on the same virtues. Thus, *pace* Emery's emphasis on theoretical virtues,[27] methodological naturalism does not *per se* warrant the inference "scientific realism → realism about metaphysics". Indeed, the fact that metaphysics relies almost exclusively on theoretical virtues is often regarded as a reason for being sceptical towards it, even in a naturalistic setting. We will not further discuss this here.[28] But what is the ground for scientific realism, then? And can it not be extended to metaphysics anyway? Let us see.

---

[26] See, *e.g.*, R. ARROYO, J. R. B. ARENHART, "The Epistemic value of Metaphysics", *Synthese*, 200(337), 2022, pp. 1–22.

The customary argument (or perhaps it is just an intuition) for scientific realism is the 'no miracle' argument, according to which the enormous success of science would be a miracle if scientific theories were not at least approximately true. This is already something that hardly applies to metaphysics, as the success being referred to is empirical success, *i.e.*, the ability to account for observed data, lead to repeatable experimental tests and enable practical and technological applications. All things that (it is easy to see) metaphysicians cannot claim to have provided. What is more, most if not all contemporary participants in the scientific realism debate acknowledge that the 'no miracle' argument/intuition can hardly convert the sceptic and something more is required. The best candidate in this connection is arguably the ability to make successful unexpected new predictions.[29] This is a feature that some successful theories lack, and for this reason alone, at least according to scientific realists who rely on novel predictions, such theories may be better looked at from an instrumentalist viewpoint. *A fortiori*, this is a feature that, methodological continuity notwithstanding, metaphysical theories lack — almost by definition. Hence realism about metaphysics is not warranted.

In more general terms, it looks as though for any scientific theory T that made successful novel predictions, or possesses any other feature of science that one can take to be at least an indicator that T is on the right track, whatever metaphysical gloss M we attach to T will not itself lead to further novel predictions. Consequently, the best argument for scientific realism does not carry over to realism about metaphysics. It may be objected that those scientific theories for which realist commitment appears warranted are in any case preliminarily selected based on theoretical virtues, hence theoretical virtues are in fact truth-conducive. We don't think is the case. For, scientists simply never find themselves in the position of choosing between empirically equivalent alternatives based on extra-empirical factors. Moreover, whatever assessment of the virtues of the 'winning theory' one may make in comparison to the 'losing' alternatives, it is typically made post hoc, the theoretical virtues of the theory that has come to be accepted by the community being more highly regarded than those of the others, without anything like an overarching ranking of virtues.

---

[27] N. EMERY, *Naturalism Beyond the Limits of Science: How Scientific Methodology Can and Should Shape Philosophical Theorizing*, New York, Oxford University Press, 2023, Chap. 3.

[28] See J. LADYMAN, "Science, Metaphysics and Method", *Philosophical Studies*, 160, 2012, pp. 31–51 and J. SAATSI, "Explanation and Explanationism in Science and Metaphysics", in M. SLATER, Z. YUDELL (eds.), *Metaphysics and the Philosophy of Science: New Essays*, Oxford, Oxford University Press, 2017, pp. 163–192.

[29] See M. ALAI, "The Historical Challenge to Realism and Essential Deployment", in T. LYON & P. VICKERS (eds.), *Contemporary Scientific Realism: The Challenge from the History of Science*, Oxford, Oxford University Press, 2001, pp. 183–215.

In light of the foregoing, clear limits emerge of the methodological continuity that the naturalist can posit: methodology is the same in science and in metaphysics *up to the point* where the former, but not the latter, proves able to provide new predictions, gain access to unobservable entities via causal interactions *etc.* — which is where science and metaphysics diverge (again). To be clear, we don't claim that realism about metaphysics is not an option. Our point is, rather, that compelling explicit arguments need to be provided in its support, other than the mere reference to naturalistic methodology. Lacking those, one can drop the parity thesis. If this is done, the option emerges of defending a non-eliminativist, yet non-realist approach to metaphysics, regardless of one's epistemic attitude towards science. Here, of course, fictionalism about metametaphysics becomes relevant.

## The benefits of fiction

Having provided reasons for dropping the parity thesis and taking fictionalist metametaphysics seriously (even in a naturalistic setting), we close by briefly mentioning (some of) the advantages we take it to entail and looking at a case study. Besides enriching the spectrum of positions about science and metaphysics that are available, in particular, we take fictionalism to bring with itself the following advantages:

> *(i) Minimal notion of truth.* As argued, fictionalism makes justice to the traditional idea that metaphysics deals with the fundamental features of reality. However, it does so by detaching the role that this idea plays in the practice of metaphysics from the notion of objective, capital-T, external truth involved in a general realistic perspective on metaphysics itself.
>
> *(ii) Better grounds for the use of theoretical virtues.* With Truth out of the way, one may use theoretical virtues freely. Once the rules are specified (*e.g.* is there a hierarchy of theoretical virtues? If so, weigh the virtues accordingly; if not, just pick the virtues you prefer and make your choice explicit), one may use theoretical virtues 'within the game'. In any case, realism having been bracketed, there is no need to search for the one correct algorithm for the evaluation of theoretical virtues nor, of course, to regard such virtues as truth-conducive.
>
> *(iii) Better grounds for metaphysical theorising.* In a sense, fictionalism entails that ontology becomes acceptable as it is reduced to internal questions-only in something like Carnap's sense. Only ontological commitment to stuff/entities within-the-game are considered, and it makes no sense to ask what is 'truly the case'. This means that, in a fictionalist setting, the study of the most fundamental aspects of reality doesn't need a realist semantics at the meta-level. In fact, paradigmatically metaphysical questions may be pursued meaningfully without the need to settle supposedly factual truths, and only referring to facts and truths in the world as it would be were the proposed fiction true. All this, it seems to us, justifies the sort of 'freedom' that metaphysicians typically exhibit when putting forward and discussing their theories, while at the same time presenting them as being about fundamental reality.[30]

(*iv*) *Neutrality with respect to scientific realism*: while traditional realism about metaphysics seems to presuppose scientific realism (at least in a naturalistic context), and scepticism towards metaphysics naturally emerges from (naturalistic) antirealistic stances with respect to the unobservable, now a 'mixed' view turns out to be plausible, or at least available. Naturalism does not require the parity thesis but just, more plausibly, that metaphysics not be trusted more, as it were, than science.

For the time being, the above considerations must suffice. We hope that we have said enough to prevent objections to the effect that what we are proposing is not metaphysics and/or is not naturalistic enough. In a sense, we believe exactly the opposite, as the strongest grounds for taking fictionalism about metaphysics seriously — at least for us — come exactly from a careful consideration of the similarities and dissimilarities between science and metaphysics.

In connection to this, a legitimate objection to metaphysical fictionalism is that metaphysics should be understood as foundational, *i.e.*, as seeking fundamental truths about reality,[31] hence talk of 'truth within the fiction of a particular theory' is a non-starter almost by definition. To answer such a worry, we can only reiterate our previous point based on an analogy with the attitude of the practising scientist dealing with particles in a lab, and point out that the very notion of something being a fundamental entity or process may well just be relative rather than absolute, as it is the content of a fiction. More precisely, exactly in the same way in which one may utter the sentence 'Electron x hit atomic nucleus y' and take it literally but at the same time realise that they fail to have compelling arguments in favour of realism about electrons, the same may hold for metaphysical conjectures. Metaphysical structure, just as ontological content, might be understood as a placeholder that differs from fiction to fiction — from game to game, as we have it — and need not be intended in the absolute sense of traditional realism. Notice that this is *not* to say that there are no objective facts about reality, nor that no fiction can in fact correspond to reality. This characteristic distances us from recent proposals such as quantum fictivism,[32] which states that quantum ontologies do not refer to physical entities, but to fictional ones. The metametaphysical fictionalism we propose here is, on the contrary, an epistemic attitude towards metaphysics, and makes no claim about the (non-)physicality of quantum entities. The point is about the reasons that we have for believing that something is the case, and in what sense.[33]

---

[30] At the same time, we repeat, we do not subscribe to the Carnapian views concerning the meaningless of external questions, and certainly regard the comparative assessment of distinct frameworks possible and indeed philosophically relevant in typical cases.
[31] For discussion, see T. OBERLE, "Metaphysical Foundationalism: Consensus and Controversy", *American Philosophical Quarterly*, 59(1), 2022, pp. 104–6, and references therein.
[32] V. MATARESE, "Quantum Fictivism", *European Journal for Philosophy of Science*, 14(38), 2024, pp. 1–27.

Before closing, in order to provide an illustration of metametaphysical fictionalism and at the same time lend support to some of our claims in the previous sections, let us briefly look, as promised, at a specific case study in the metaphysics of science. It concerns exactly the issue concerning ontology of non-relativistic quantum mechanics we have just referred to.

Debates on quantum foundations surely testify to the difficulty of settling metaphysical matters in a clear, uncontroversial manner. Here, we will consider the sort of underdetermination that arises when one attempts to add a metaphysical layer to the physical theory, in the form of an answer to the question 'What sort of entities are those described by quantum mechanics?'.

At a first level, underdetermination arises when one deals with the measurement problem that notoriously affects quantum mechanics.[34] As it is well known, the problem consists of a tension between the theory's predictions via its mathematical framework, on the one hand, and the theory's empirical findings, on the other hand. This tension might be roughly exemplified by the fact that quantum mechanics describe physical systems that a) typically lack determinate values for their properties and evolve according to a deterministic equation (the Schrödinger equation) that preserves such property indeterminateness; and yet b) are always observed as possessing determinate properties. More precisely, in familiar quantum-mechanical situations, a physical system might have the property of, say, being located at two disjoint spatiotemporal regions, A and B; the quantum formalism then describes such a situation as the system being in a state of *superposition* of the states 'being-detected-at-A' plus 'being-detected-at-B'. This superposition situation is one of the greatest mysteries of quantum mechanics, with no classical analogue, as it cannot be interpreted as the system having the property of being located in A or B and we just happen to lack the knowledge of which is the case; in fact, the most precise thing to state is that there is no fact of the matter about the location of the physical object in question. Yet, every time the position is measured, the physical system will always be found either in A or B (with a given probability for each corresponding case). To account for an explanation of how and why such a change takes place is what counts as a solution to the measurement problem.

Ways of dealing with the measurement problem abound and differ quite radically among them. For instance, Bohmian mechanics presupposes a quasi-classical ontology of particles following determinate trajectories and possessing definite properties at all times; spontaneous collapse theories postulate two novel constants of nature determining the instantaneous shift from an indeterminate to a determinate state; many worlds interpretations conjecture a multiplicity of fully determinate realities, so that each one of the possible states of a system is realised in one of them; relational quantum mechanics makes the possession of a determinate property by a physical system possible insofar as it is always relative to another system, and so on. The point to emphasise here is that the options are underdetermined concerning the empirical data.

---

[33] We thank an anonymous referee pressing us on this issue.
[34] See M. Egg, J. Saatsi, "Scientific Realism and Underdetermination in Quantum Theory", *Philosophy Compass*, 16(1), p. e12773, 2021.

In addition to this kind of underdetermination of the theory by the empirical facts, there are others. At a second level, underdetermination arises when one asks what sort of entities the theory, in one of its formulations/interpretations, describes. Candidates include particles, fields, relational structure, flashes, waves in configuration space and more. At a third level, one can ask further questions about the metaphysical nature of the postulated entities. Suppose, for instance, that we take quantum theory to be about particles: what sort of objects are they? What are their identity conditions?

Let us look at this specific set of issues more closely. On the one hand, unlike classical objects, quantum objects may be permuted without this making any difference with respect to statistics, *i.e.*, to the computation of the number of states available for a system of many particles. Consider the case in which two particles (call them '1' and '2') are to be distributed in two disjoint spatial regions A and B. There are exactly four ways of doing so: i) both 1 and 2 in region A, or 'A(1,2)'; both 1 and 2 in region B, or 'B(1,2)'; iii) particle 1 in region A and particle 2 in region B, or A(1)B(2); iv) particle 2 in region A and particle 1 in region B, or A(2)B(1). This is the classical or 'Maxwell–Boltzmann' statistics and each of the four possible physical arrangements has the same statistical weight. Classical particles obey such statistics. This is not the case for quantum particles. Call a quantum particle '•'. In symbols, we have just three cases: A(••), B(••), and A(•)B(•). Classical cases A(1)B(2) and A(2)B(1) are collapsed into quantum case A(•)B(•), as a 'permutation', *i.e.*, a switch of one particle with another, does not generate a different physical situation. This is called 'Bose–Einstein' statistics, or quantum statistics. All that matters in the quantum scenario is the cardinality of the three cases: there are exactly two particles in region A, two particles in B, or exactly one particle in each region A and B. In the conventional interpretation of this situation, it's believed that, due to this statistical reason, while classical entities possess individuality, quantum entities do not. This is what is known as the Received View of quantum (non-)individuality:[35] The shift in statistics, to reiterate, is due to a change in the metaphysical nature of quantum entities. Another important fact about quantum entities is that they seem to violate the principle of the Identity of the Indiscernibles, at least in its traditional formulation according to which two numerically distinct entities differ with respect to at least one of their monadic qualitative properties. If this is the case, non-individuality may be used to explain the exact qualitative similarity of 'two different things'[36].

---

[35] J. R. B. ARENHART, "The Received View on Quantum Non-individuality: Formal and Metaphysical Analysis", *Synthese*, 194(4), 2017, pp. 1323–47.
[36] The debate, however, is quite complex, and there are several other options. See T. BIGAJ, *Identity and Indiscernibility in Quantum Mechanics*, Palgrave-Macmillan, 2022.

But there's a twist, as quantum objects may also count as individuals: one may always interpret them as having a non-qualitative property of 'being identical with itself', framing such 'thisness' in terms of substances, non-qualitative *haecceitates* or as primitive features.[37] The fact that quantum individuals obey a statistics which is different from classical statistics would then be explained on the basis of hypotheses concerning not the identity conditions of quantum objects but rather their properties. Attributing such a metaphysical profile of individuality to (quantum or else) physical entities is, as Esfeld puts it, a "[…] a purely metaphysical move that one can always make, physics be as it may".[38]

Faced with this, one may deny that truth is an aim of scientific enquiry — either uniquely in the quantum domain,[39] or for any scientific endeavour.[40] *A fortiori*, when it comes to metaphysics, one could follow van Fraassen in believing that, because we cannot settle these metaphysical matters, we should say "good-bye to metaphysics".[41] On the other side, scientific-cum-metaphysical realists typically attempt to break the relevant underdetermination by having recourse to theoretical virtues.[42] As we have already mentioned, however, theoretical virtues do not seem to be truth-conducive. In some cases, the attempt is made to sidestep underdetermination by finding something shared by all the underdetermined alternatives. There are reasons, however, for thinking that attempts to break underdetermination in this way in fact lead to further underdetermination. For instance, structuralists about quantum entities claim that positing an ontology of structures sidesteps the individuality issue entirely. Yet, arguably structuralism can itself be formulated in many ways, and such a structural underdetermination may be taken to strike a fatal blow to the structuralist strategy in the quantum domain.[43] Incidentally, notice that underdetermination may extend to the basic formal tools that we take to be instrumental to describing the relevant physical domain, e.g., to the choice between 'non-classical' quasi-set theory and 'classical' Zermelo–Fraenkel set theory for quantum mechanics.[44]

Faced with this kind of situation, what should we do? *Prima facie*, it looks as though we are led back to what we presented earlier as the parity thesis: either antirealism about both science and metaphysics (possibly with elimination of the latter), or scientific realism accompanied with the hope that somehow we will get at the correct metaphysical framework for our best science — in this case, non-relativistic quantum mechanics.

---

[37] M. MORGANTI, "The Metaphysics of Individuality and the Sciences", *in* T. PRADEU, A. GUAY, (eds.), *Individuals Across the Sciences*, Oxford, Oxford University Press, 2015, pp. 273–94.

[38] M. ESFELD, "Ontic Structural Realism and the Interpretation of Quantum Mechanics", *European Journal for the Philosophy of Science*, 3, 2013, pp. 19–32.

[39] C. HOEFER, "Scientific Realism without the Quantum", *in* J. SAATSI & S. FRENCH (eds.), *Scientific Realism and the Quantum*, Oxford, Oxford University Press, 2020, pp. 19–34.

[40] B. VAN FRAASSEN, *Quantum Mechanics: An Empiricist View*, Oxford, Oxford University Press, 1991.

[41] *Ibid.*, p. 480.

[42] C. CALLENDER, "Can We Quarantine the Quantum Blight?", *in* J. SAATSI & S. FRENCH (eds.), *Scientific Realism and the Quantum*, Oxford, Oxford University Press, 2020, pp. 57–77.

[43] See O. BUENO, "Revising Logics", *in* J.-Y. Béziau. *et al.* (eds.), *Logic in Question: Talks from the Annual Sorbonne Logic Workshop (2011–2019)*, Cham, Springer, 2022, pp. 303–20.

[44] On this, see R. ARROYO, J. R. B. ARENHART, D. KRAUSE, "The Elimination of Metaphysics through the Epistemological Analysis: Lessons (un)Learned from Metaphysical Underdetermination", *in* D. AERTS *et al.* (eds.), *Probing the Meaning of Quantum Mechanics: Probability, Metaphysics, Explanation and Measurement*, Singapore, World Scientific, 2023, pp. 278–324.

Fictionalism, however, offers a nice middle ground: on the one hand, it makes justice to the very attempt to find precise answers to metaphysical questions; on the other, it refrains from thinking that there must be one correct answer. In a fictionalist context, for instance, one can decide to work with a 'conservative' ontology and consequently pick, say Bohmian mechanics with an ontology of individual particles. But one can equally decide to start with spontaneous collapse interpretations of quantum mechanics as a promising solution to the measurement problem and, on that basis, develop a consistent ontology evaluating the various options with reference to theoretical virtues (in addition of course to the empirical data). To repeat, this is entirely compatible with a scientific realist approach to the relevant theory and even with a particular metaphysical gloss on it. The relevant element in a fictionalist context is just that one can trust a hypothesis (be it scientific or metaphysical) even if they just accept it as a tool of inquiry and a source of explanation, and feel that — as things stand at a given time — they have no good reasons (yet!) for believing it, *i.e.*, to believe that it provides an approximately true description of the relevant part of reality.

**Acknowledgements**

Earlier versions of this article were presented on several occasions: the X International Workshop on Quantum Mechanics and Quantum Information, held at Roma Tre University, 2024; the Foundations/Philosophy of Physics Seminar from Utrecht University, 2024; the IV Colloquium of the Metaphysics Working Group, held at Puc-Rio, 2024; and the "Philosopher's reply" workshop, held at the Federal University of Santa Catarina, 2024. As such, its final version benefited from its participants: Adonai Sant'Anna, Andrea Oldofredi, Bruno Borge, Carl Hoefer, Cristian López, Davide Romano, Décio Krause, Eliza Wajch, Emanuele Rossanese, Emilia Margoni, Enrico Cinti, F. A. Muller, Francesco Ferrari, Jonas Arenhart, Mauro Dorato, Olimpia Lombardi, Otávio Bueno, Pedro Merlussi, Samuele Fasol, Valia Allori, and Vera Matarese.

**Funding**

Raoni Arroyo was supported by grant #2021/11381-1, São Paulo Research Foundation (FAPESP). This article was produced during his research visit to the Department of Philosophy, Communication, and Performing Arts, Roma Tre University, Rome, Italy, supported by grant #2022/15992-8, São Paulo Research Foundation (FAPESP). Matteo Morganti was supported by PRIN 2022, "The Philosophical Reception of Quantum Theory in France and German-speaking Countries between 1925 and 1945: Conceptual Implications for the Contemporary Debate", prot. 20224HXFLY.